\documentclass[final,5p,times,twocolumn]{elsarticle}

\usepackage{amssymb}
\usepackage{amsmath}

\graphicspath{{figures/}}
\usepackage{siunitx}

\journal{International Journal of Mass Spectrometry}

\usepackage{xspace}
\newcommand{\ba}{$^{138}$Ba$^+$\xspace}
\newcommand{\ca}{$^{40}$Ca$^+$\xspace}

\newcommand{\harf}{\mbox{$\frac12$}}

\newcommand{\aeff}{a_{\mathrm{eff} } }

\newcommand{\Vppm}{V^{\prime\prime}_{\mathrm{max}}}

\newcommand{\VL}{V_{1} }
\newcommand{\Fp}{F^{\prime} }

\newcommand{\qtau}{ q^{\{\tau \} }_A }

\newcommand{\qt}{ q^{(\tau_n)}_A }
\newcommand{\aplus}{a^+}
\newcommand{\aminus}{a^-}
\newcommand{\kAz}{\kappa_{A, z} }
\newcommand{\Vz}{V_{0}^{ (z) } }
\newcommand{\kBz}{\kappa_{B,z} }

\newcommand{\Om}{\Omega}
\newcommand{\omA}{\omega_A}

\newcommand{\OmL}{\Omega}
\newcommand{\OmH}{\Omega_{n}}

\newcommand{\reffig}[1]{Fig.~\ref{#1}}
\newcommand{\refeq}[1]{Eq.~\ref{#1}}
\newcommand{\reftable}[1]{Table~\ref{#1}}
\newcommand{\refsec}[1]{Section~\ref{#1}}

\newcommand*{\GtrSim}{\smallrel\gtrsim}

\makeatletter
\newcommand*{\smallrel}[2][.8]{%
  \mathrel{\mathpalette{\smallrel@{#1}}{#2}}%
}
\newcommand*{\smallrel@}[3]{%
  \sbox0{$#2\vcenter{}$}%
  \dimen@=\ht0 %
  \raise\dimen@\hbox{%
    \scalebox{#1}{%
      \raise-\dimen@\hbox{$#2#3\m@th$}%
    }%
  }%
}
\makeatother

\begin{document}

\begin{frontmatter}



\title{Two-frequency operation of a Paul trap to optimise confinement of two species of ions}


\author[ox]{C. J. Foot\corref{cor1}}
\ead{c.foot@physics.ox.ac.uk}
\author[jqi]{D. Trypogeorgos}
\author[ox]{E. Bentine}
\author[sus]{A. Gardner}
\author[sus]{M. Keller}

\address[ox]{Clarendon Laboratory, Department of Physics, University of Oxford, Parks Road, Oxford, OX1 3PU, UK}
\address[jqi]{Joint Quantum Institute, University of Maryland and National Institute of Standards and Technology, College Park, Maryland, 20742, USA}
\address[sus]{Department of Physics and Astronomy, University of Sussex, Falmer, BN1 9QH, U.K.}

\cortext[cor1]{Corresponding author}

\begin{abstract}
We describe the operation of an electrodynamic ion trap in which the electric quadrupole field oscillates at two frequencies.
This mode of operation allows simultaneous tight confinement of ions with extremely different charge-to-mass ratios, e.g., singly ionised atomic ions together with multiply charged nanoparticles.
We derive the stability conditions for two-frequency operation from asymptotic properties of the solutions of the Mathieu equation and give a general treatment of the effect of damping on parametric resonances.
Two-frequency operation is effective when the two species' mass ratios and charge ratios are sufficiently large, and further when the frequencies required to optimally trap each species are widely separated.
This system resembles two coincident Paul traps, each operating close to a frequency optimized for one of the species, such that both species are tightly confined.
This method of operation provides an advantage over single-frequency Paul traps, in which the more weakly confined species forms a sheath around a central core of tightly confined ions.
We verify these ideas using numerical simulations and by measuring the parametric heating induced in experiments by the additional driving frequency.
\end{abstract}


\end{frontmatter}

The Paul trap~\cite{paul_electromagnetic_1990-1,paul_electromagnetic_1990} confines charged particles using an oscillating quadrupole electric field thus circumventing Earnshaw's theorem for static fields~\cite{earnshaw_nature_1848}.
This highly versatile method of electrodynamic confinement has a multitude of applications, spanning a wide range of charge-to-mass ratios $Q/M$, which may be broadly categorised by the drive frequency of the trap.
For example, guiding of electrons by microwave fields was recently achieved~\cite{hoffrogge_planar_2011}, but practical difficulties of confining them in a Paul trap include driving frequencies of the order of \SI{1}{GHz}.
The majority of current experiments to confine atomic ions (\ba, \ca,  etc.) use a quadrupole electric field oscillating in the range of tens of MHz.
Commercial mass spectrometers often have stages which guide ions along linear quadrupole fields, operating at MHz frequencies for light ions, down to tens of kHz for larger, heavy molecules with lower $Q/M$.
Electrodynamic balances are operated at 50 or 60\,Hz to confine micron-sized particles of even lower charge-to-mass ratios~\cite{nasse_influence_2001,winter_simple_1991}.
The achievable $Q/M$ typically decreases as the mass of the ion increases~\cite{huang_gas-phase_2010,schiller_molecular_2003,hilton_two_2012}, shifting optimal parameters towards lower drive frequencies.

The dependence of the Paul trap on the mass and charge of the ions makes it difficult to confine two species for which the optimal driving frequencies are very different; typically the heavier species has a lower $Q/M$ ratio leading to weaker confinement by the oscillating quadrupole field.
As a result, the interaction between the two species is proportionally weaker.
This hinders the use of fluorescing light ions as either a detector for the heavier dark ions, or for sympathetic cooling~\cite{ostendorf_sympathetic_2006}.
For similar charge-to-mass ratios a third intermediate species can be used to bridge the gap between the first two~\cite{zhang_molecular-dynamics_2007}, but this becomes impractical for extremely dissimilar ratios.
A major obstacle in pioneering work on the simultaneous trapping of atomic \ba and molecular ions with mass \SI{410}{Da} in a linear Paul trap operating at a single frequency~\cite{offenberg_translational_2008}, was the spatial separation of the constituents that occurred because of their mismatched spring constants.
Ions of the more weakly confined species were pushed away from the trap centre by Coulomb repulsion from the strongly trapped ions, leading to a sheath of molecular ions around a core of the more tightly confined \ba.
For mixtures of more dissimilar ions, the Coulomb repulsion of the core may even prevent trapping of the heavier ions.

Here we propose an approach motivated by an aspect of electrodynamic trapping --- hinted at by Dehmelt in a one-page bulletin on confinement of antimatter~\cite{dehmelt_economic_1995} --- that has received little attention.
Dehmelt suggested using electric fields oscillating at two suitably chosen frequencies to simultaneously confine charged particles with different $Q/M$; this idea has recently been explored further in~\cite{leefer_investigation_2016}.
We show that the same principle can be used to confine atomic ions and heavier charged molecules.
It is straightforward to adapt a standard Paul trap apparatus for atomic ions so that it also confines charged particles of higher mass (lower charge-to-mass ratio than atomic ions) using two driving frequencies.
The two-frequency trap exploits the implicit link between $Q/M$ of the charged particles and the frequency for optimal operation of a Paul trap.

Two-frequency operation of a Paul trap offers a significant advantage when confining two species with widely different properties, corresponding to widely different optimal drive frequencies~\cite{trypogeorgos_cotrapping_2016}.
It increases the strength of confinement of the heavier species thus making the spring constants more similar and bringing the two species closer together.
This in turn increases the interaction between the otherwise well separated ion species.
Here, we solve the system of equations that describes the stability of a single charged particle in a quadrupole electric field oscillating at two frequencies, and interpret the solutions in terms of the pseudopotential approximation that gives an intuitive picture in the single-frequency case.
By considering parametric resonances that determine the stability of this system, we identify threshold criteria for stable two-frequency operation with two species.
High-order parametric resonances are sensitive to damping of the motion~\cite{nasse_influence_2001,hasegawa_dynamics_1995,pedersen_stability_1980}, as we demonstrate experimentally, and they can be suppressed, e.g.\ by laser cooling or a buffer gas.

This paper is organised as follows.
We start by reviewing the theory of the Paul trap in \refsec{sec:paul} and consider the spring constants for two trapped species.
The theory of operation of the two-frequency trap is developed in \refsec{sec:twofreq}.
Parametric resonance sets a limit to how close the charge-to-mass ratios of two trapped ions can be when seeking to enhance the spring constant of weakly trapped species.
In \refsec{sec:exp} we present parametric heating measurements of a single trapped ion in a two-frequency trap and compare the results with our theoretical model.
These results are extended further using molecular dynamics (MD) simulations in \refsec{sec:md}.
Our numerical results show how the effect of increasing the spring constant of a weakly trapped, dark, heavy ion can be observed through the deformation of a fluorescing atomic ion cloud.
We conclude with a discussion of potential applications of our proposed method in \refsec{sec:concl}.

\section{Theory of the Paul trap}
\label{sec:paul}

Newton's equation of motion for a particle of mass $M$ and electric charge $Q$ along the $y$-axis of a Paul trap is $M(\ddot{y} + \Gamma\dot{y}) = Q\,E_y(t)$, where $\Gamma$ is a damping constant.
The time-dependent force equals the charge $Q$ times the component of the electric field $E_y$ along this axis.
The quadrupole field is $E_y(t) = V(t)\,y/r_0^2$, where the distance $r_0$ characterises the electrodes' spacing, and the applied voltage is $V(t) = V_0 + V\cos(\Omega t)$.
A change of variables $\Omega t = 2\tau$ transforms the equation of motion into the Mathieu equation~\cite{paul_electromagnetic_1990-1, berkeland_minimization_1998, wineland_experimental_1998, ghosh_ion_1996}
\begin{equation}\label{eq:Mathieu}
\frac{d^2 y}{d\tau^2} + \left[ a- 2q\cos(2\tau)\right] y(\tau)=0,
\end{equation}
with $a = (Q/M\Omega^2)\cdot 4 V_0/r_0^2$ and $q  = -(Q/M\Omega^2)\cdot 2 V / r_0^2$.
An approximate method that elucidates the behaviour of the Paul trap for certain parameters shows that the motion of the ion consists of an oscillation at a slow secular frequency plus a fast, small-amplitude micromotion at the driving frequency: $y= A\cos(\omega t)(1 + \frac 12 q \cos(\Omega t))$.
The slow motion is that of a particle in a harmonic pseudopotential with secular frequency $\omega$ given by  the Dehmelt approximation
\begin{equation}\label{eq:omegaaq}
\omega = \frac{\Om}{2} \sqrt{a + \frac 12 q^2}.
\end{equation}
More generally, the motion of charged particles in an oscillating electric field may be described by a ponderomotive potential that is proportional to the square of the amplitude of the oscillating electric field.
\begin{figure}[t]
\centering
\includegraphics[]{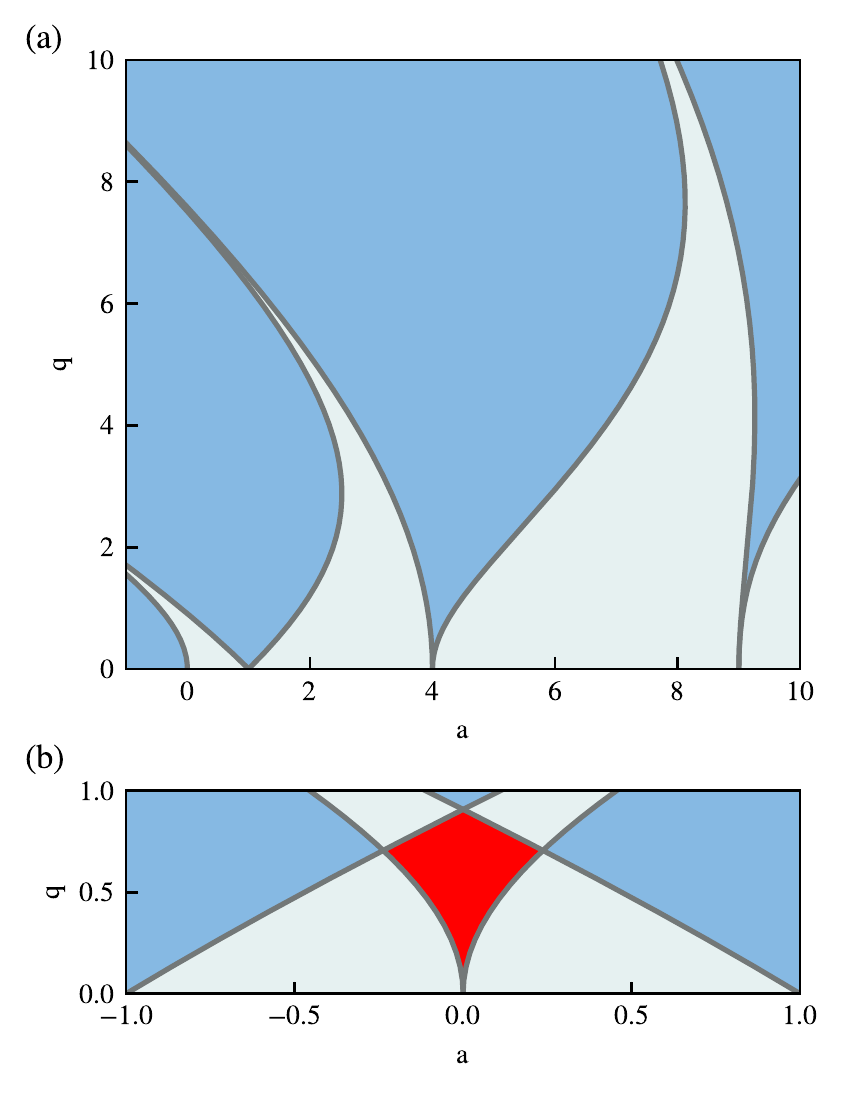}
\caption{(a) Ince-Strutt stability diagram for the Mathieu equation (\refeq{eq:Mathieu}).
The solutions are stable and periodic in the lightly coloured areas and diverging otherwise.
(b) The stability region for a linear Paul trap, shown in red,  is the overlap of the stability region for motion along the $x-$ and $y-$ axes. A linear quadrupole field has reflection symmetry along the line $a=0$ since $a_y=-a_x$.}
\label{fig:1}
\end{figure}
The stability region of the one-dimensional Mathieu equation (\refeq{eq:Mathieu}) is shown in \reffig{fig:1}(a).
Stability in both $x$- and $y$-axes is independent of the sign of $q$, and hence the sign of the charge $Q$.
The Paul trap is stable when $q>0$ even for negative values of $a$ since a small amount of anti-trapping by the static electric field can be overcome by the pseudopotential, e.g., for $q=0.4$ stable solutions exist for $\vert a \vert < 0.09$.
Figure~\ref{fig:1}(b) shows the intersection of the stability regions for motion in the $x$ and $y$ directions of a linear Paul; values of the parameters $a$ and $q$ within this overlap region give stable confinement.
In a linear Paul trap, the motion of the ion is stable over the range $0 < q < 0.91$ for $a=0$.

Values of $q \le 0.4$ give robust operation of a Paul trap without anharmonic effects that tend to destabilise the confinement at higher values.
Therefore we use $q = 0.4$ as a typical value throughout this paper, although the results do not depend on the exact value chosen.
Practical considerations limit the attainable curvature $V^{\prime\prime}_{max} =V/r_0^2$ for a given trap.
Realistic values are $V = \SI{500}{V}$ and $r_0 = \SI{0.5}{mm}$ giving $V^{\prime\prime}_{max} \simeq \SI{2e9}{V/m^2}$ which is comparable to the value used in the measurements described in \refsec{sec:exp}.
This choice of $V^{\prime\prime}_{max}$ determines the driving frequency that gives a suitable values of the $q$-parameter, i.e., $q=0.4$.

\subsection{Confinement of two species in a single-frequency Paul trap}

We consider two species A, B trapped in a Paul trap as in our previous theoretical work~\cite{trypogeorgos_cotrapping_2016}; species A is atomic \ba and B a heavy ion with $M_B = \SI{1.4e6}{Da}$ and $Q_B=+33e$.
The spring constant $\kappa = M \omega^2$ determines the extent of the ion cloud $y_\mathrm{rms}$ at temperature $T$, and hence the overlap of the two species, since by the equipartition theorem $k_B T =\kappa y_{\mathrm{rms} }^2$.
At low $T$ the ions form Coulomb crystals and electrostatic repulsion determines their spatial extent: equal and opposite trapping forces between two ions $\kappa_A y_A = -\kappa_B y_B$ result in similar displacements, $\vert y_A \vert \approx \vert y_B\vert $, if the spring constants are similar $\kappa_A \approx \kappa_B$.
For most of these calculations it is not necessary to consider the Coulomb interaction of the trapped ions, however it is included in the numerical simulations of \refsec{sec:md}.
In the pseudopotential approximation the spring constant is
\begin{equation}\label{kappalong}
\kappa \approx M \omega^2 = \frac{q}{8}\cdot \frac{2 Q V}{r_0^2} = \frac{Q^2} {M\Omega^2}\cdot\left(\frac{V}{2 r_0^2} \right)^2,
\end{equation}
where $\omega \simeq q \Omega /\sqrt{8}$ when $a=0$.
The ratio of the two spring constants for the single-frequency trap is:
\begin{equation}\label{eq:KBKA}
\left[\frac{ \kappa_B }{ \kappa_A}\right]_\mathrm{1rf} = \frac{ \kappa_B(\Omega) }{ \kappa_A(\Omega)} = \frac{ Q_B^2/ M_B }{ Q_A^2/M_A },
\end{equation}
where $\kappa_A(\Omega)$, $\kappa_B(\Omega)$ are both functions of $\Omega$.
For the species considered here $(Q_B^2/ M_B)/(Q_A^2/M_A)=0.1$, so that the confinement of species B is less tight; these ions move to the outside of the cloud of species A in a single-frequency trap since $q \propto Q/M$ for fixed $\Omega$, and $q_A =0.4$ for species A implies $q_B = 0.001$.

The spring constant of a Paul trap operating at a fixed voltage $V$, and with drive frequency optimized such that $q=0.4$, depends only on the ion's charge (see \refeq{kappalong}).
If it were possible to simultaneously achieve the same conditions for both species, the ratio of their spring constants would be $\kappa_B/\kappa_A = Q_B/Q_A \gg 1$,
which is opposite to normal for a Paul trap (see \refeq{eq:KBKA}).
However, this scenario cannot be attained in a single frequency trap.
The constraints that $q = 0.4$ for fixed $V$ are incompatible for the two species, each requiring a different drive frequency.
Much of this work is devoted to optimisation of the confinement of species B, to give the maximum $\kappa_B$, subject to the constraint that species A remains trapped.
We show that two-frequency operation gives significant improvement over a single-frequency Paul trap in suitable circumstances.

\section{Trapping with two frequencies}
\label{sec:twofreq}

We can optimise the confinement of each species individually by using different driving frequencies: a high frequency $\OmH$ for the light ions and a low frequency $\OmL$ for the heavier ones.
Applying the second, lower frequency field augments the confinement of the heavier species.
The ratio of spring constants now becomes
\begin{equation}\label{eq:KBKALoHi}
\left[\frac{ \kappa_B }{ \kappa_A}\right]_\mathrm{2rf} = \frac{ \kappa_B(\OmL) }{ \kappa_A(\OmH)} =\left(\frac{V_1}{\Omega}\cdot\frac{\Omega_n }{V_n}\right)^2\frac{ Q_B^2/ M_B }{ Q_A^2/M_A },
\end{equation}
where $V_1$, $V_n$ are the voltages that correspond to $\OmL$, $\OmH$.
Comparison with \refeq{eq:KBKA} shows that there is an enhancement by a factor of $\left(\Omega_n/\Omega\right)^2\cdot \left(V_1/V_n\right)^2$, which can be much greater than unity as shown below.
Although $V_1 \ll V_n$ because the low-frequency component strongly influences the stability of species A, an overall enhancement can be achieved when $\Omega_n/\Omega \gg 1$ is sufficiently large.

To configure the two-frequency trap we first set the values $V_n$, $\OmH$ to be optimal for single frequency confinement of the light ion with $q_A \simeq 0.4$.
With species A now well confined, we chose parameters $\OmL$, $V_1$ to trap species B, subject to the requirement that there is no parametric excitation of A.
This approach ensures that both species are stable in the two frequency quadrupole field.

\subsection{Parametric resonance}
We consider a  system of two species A and B subjected to two frequencies $\OmL$ and $ \OmH$, where the higher frequency is a harmonic $\OmH = n \OmL$ of the lower driving frequency; the assumption that $n$ is an integer allows use of the convenient mathematics of periodic systems but is not a physical requirement.
In a standard single-frequency Paul trap the secular oscillation frequencies of the two species are $\omega_B = q/\sqrt{8}\cdot \OmL$ and $\omega_A = q/\sqrt{8}\cdot \OmH$.
These four frequencies are summarised in \reftable{tab:1}.

\begin{table}[ht]
\centering
\caption{The four relevant frequencies: $\omega_A$ is the secular frequency of ions of species A in a Paul trap driven at $\OmH$; similarly $\omega_B$ for species B with $\OmL$ only. When both driving frequencies are present the field at $\OmL$ can parametrically excite species A but the quadrupole field oscillating at $\OmH$ only weakly affects species B.}
\begin{tabular}{lrr}\\
species$\setminus$rf & $\OmH$ & $\OmL$ \\
\hline
A  & $\omega_A$ & parametric excitation \\
B  & weak trapping & $\omega_B $ \\
\hline
\end{tabular}
\label{tab:1}
\end{table}

There is a natural ordering of the frequencies:
$\omega_B < \OmL<\omega_A < \OmH$.
The first and third inequality are satisfied as a consequence of the standard single-frequency operation of a Paul trap.
The middle inequality is important for the two-frequency operation; $\OmL$ must be well below $\omega_A$ to avoid driving a parametric resonance which leads to heating of species A; this limits the amplitude of the low frequency field.

We first consider the stability of species B when the voltage $V(t) = V_0 + V_1\cos(\OmL t) + V_n\cos(n\OmL t)$ drives the quadrupole trap electrodes. We rescale time to $\OmL t = 2 \tau$ so that
\begin{equation}\label{eq:yB}
\frac{d^2 y_B}{d\tau^2} + \frac{1}{V_B}\left(V_0 + V_1 \cos(2\tau) + V_n \cos(2n\tau) \right)y_ B = 0,
\end{equation}
where $V_B = (M_B/Q_B) \cdot \OmL^2r_0^2/4$.
The dimensionless coefficients of the Mathieu equation are given here as voltage ratios.
This has the form of a Hill equation: $y^{\prime\prime} + H(\tau) y =0$, where $H(\tau) = H(\tau +\pi)$ is a periodic function\footnote{The Mathieu equation is an example of a Hill equation with only a single periodic coefficient.}.
The high frequency component oscillating at $\OmH$ has little effect on the stability of species B, which remains well described by a single-frequency trap operating at the low frequency $\OmL$.
\begin{figure}[t]
\centering
\includegraphics[]{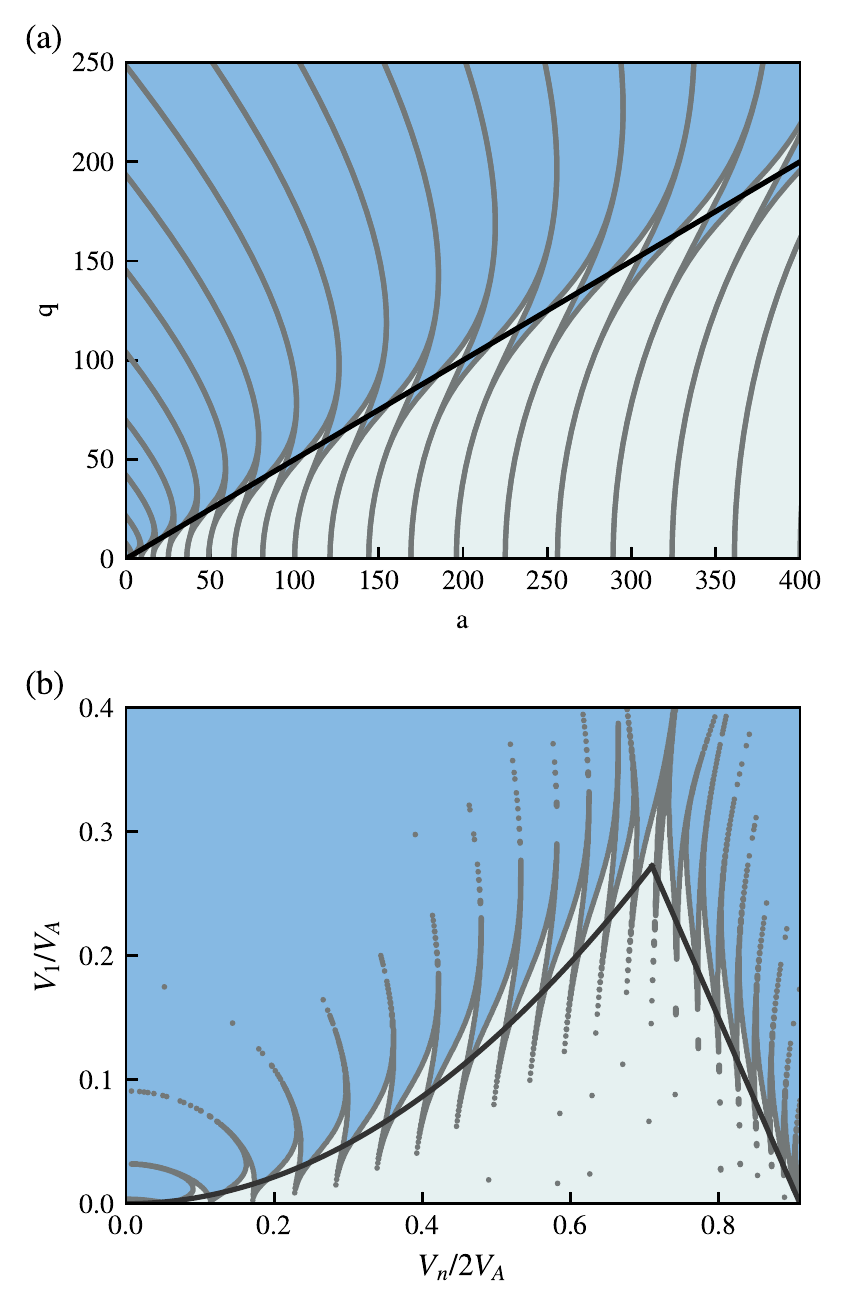}
\caption{Stability diagrams of a Paul trap for (a) single- and (b) two-frequency operation where $n=25$ and $a=0$.
The darker regions correspond to the tongues of instability which extend all the way to the horizontal axis, although this cannot be seen due to the finite precision of the numerics.
The black lines are the critical curves for both configurations.
The critical curve is $a=2q$ in the single frequency case, while in the two-frequency case it increases quadratically and then descends linearly  (see text for more details).
}
\label{fig:2}
\end{figure}

The same is not true for species A, as parametric resonances lead to heating when $m\OmL = 2 \omega_A$; here the integer $m$ is the order of the resonance with $m < n$ since $\omega_A \sim 0.14 n \OmL$.
Consider the equation of motion for species A, rescaled at the higher frequency such that $\tau_n = n \OmL t / 2 = n\tau$
\begin{equation}\label{eq:yA}
\frac{d^2 y_A}{d\tau_n^2} + 2\frac{\beta}{n}\frac{d y_A}{d\tau_n} + \frac{1}{V_A}\left(V_0 +V_1\cos\left(\frac{2\tau_n}{n}\right)+V_n\cos (2\tau_n) \right) y_A=0,
\end{equation}
where $V_A = (M_A/Q_A) \cdot n^2 \OmL^2r_0^2/4$.
Note the factor of $n^2$ compared to the definition of $V_B$ in \refeq{eq:yB}.
The scaled damping parameter $\beta = \Gamma_A/\OmL$ accounts for cooling of species A (see~\ref{sec:damp} for a simple mathematical treatment of damping).
Equations~\ref{eq:yB} and~\ref{eq:yA} are ordinary differential equations with periodic coefficients~\cite{jordan_nonlinear_2007} and can be solved using Floquet theory which considers the mapping from the solution at time $t$ to that at time $t+T$ where $T$ is the period of the system~\cite{floquet_sur_1883,konenkov_matrix_2002}.
The influence of any dc field $\propto V_0$, the Coulomb repulsion between ions, and damping can be included by standard numerical methods.

Figure~\ref{fig:2}(b) shows the stability region for an ion of species A as a function of $V_n/2V_A$ and $V_1/ V_A$ obtained by numerical solution of \refeq{eq:yA}~\cite{trypogeorgos_cotrapping_2016}.
Parametric excitation by $V_1 \cos (\OmL t)$ causes $n-1$ tongues of instability emanating from the horizontal axis $V_1=0$ which get wider as $V_1$ increases.
The tongues of instability, aka Arnold tongues~\cite{arnold_mathematical_1989}, arise for periodic systems, including the Mathieu equation, and their properties have been studied extensively for the more general case of Hill
equations~\cite{weinstein_asymptotic_1987,simakhina_computing_2005,roncaratti_whittakerhill_2010}.
For $n\GtrSim 10$ the stability regions show universal behaviour: a quadratic increase up to $V_n/2V_A \simeq 0.7$ followed by a linear decrease to zero at $V_n/2V_A = 0.91$.
The critical curve that traverses equal distances in the stable and unstable regions \cite{broer_large_2013, weinstein_asymptotic_1987} over the range $0< V_n/2V_A < 0.71$ is a quadratic curve of the form:
\begin{equation}\label{eq:critical}
\frac{V_1}{V_A} =\rho \left(\frac{V_n}{2V_A}\right)^2,
\end{equation}
where $\rho=0.54$ is a constant.
This critical curve has a different functional form than the single frequency case $a=2q$, shown in \reffig{fig:2}(a), yet it derives from a similar argument.
We rewrite \refeq{eq:yA} so that it resembles the Mathieu equation, assuming $V_0 = \beta = 0$
\begin{equation}\label{eq:yAsimple}
\frac{d^2 y_A}{d\tau_n^2} + \left[  \alpha(\tau_n) - 2 \qt \cos 2\tau_n \right] y_A=0,
\end{equation}
where $- \qt = V_n /2V_A$, and $\alpha(\tau_n) = V_1/V_A$ resembles an $a$-parameter that arises from a slowly varying dc field.
Neglecting the slow variation of $\alpha(\tau_n)$, we use the condition for stability $\vert \alpha \vert < q^2/2$; this corresponds to the pseudopotential approximation and gives \refeq{eq:critical} with $\rho= 0.5$.
See~\ref{sec:damp} for a more rigorous treatment based on properties of the Mathieu equation for large $a$ and $q$.

The tongues of instability become narrower below the critical line and higher-order resonances are too fine for the numeric calculations to capture --- the finite resolution of the numerics acts as an effective damping.
The asymptotic properties of the Mathieu equation can be used to find the endpoints of the tongues for a given damping (\ref{sec:damppe}); they lie on the curve
\begin{equation}\label{eq:damping}
\frac{V_1}{V_A} =\rho \left(\frac{V_n}{2V_A}\right)^2  \left(\pi\beta \right)^{1/m}.
\end{equation}
The integer $m$ is the order of the parametric excitation equal to the number of resonances up to the given value of $V_n/ 2V_A$.
The dependence on $\beta^{1/m}$ is typical for the threshold value of resonant parametric excitation~\cite{landau_mechanics:_1976}.
The order number $m$ at a resonance is
\begin{align}\label{eq:mest}
 m = \frac{2\omA}{\OmL}  \simeq  n\frac{q}{\sqrt{2}} =\frac{n}{\sqrt{2}} \left(\frac{V_n}{2V_A}\right),
\end{align}
hence we can approximate $m \simeq 0.28n$ for $q=0.4$.
Equation~\ref{eq:damping} gives realistic values for low-order resonances~\cite{razvi_fractional_1998,zhao_parametric_2002}.
The width of the instability tongues decreases rapidly with increasing order of the parametric resonance.
For the Mathieu equation this width is proportional to $q^m$ for a resonance emanating from $(q, a) = (0, m^2)$.
Asymptotic approximations work well for narrow resonances in the limit of large $n$ as described in~\ref{sec:damppe}, whereas numerical calculations require a fine grid for such narrow features and hence long computation times.
Equations~\ref{eq:critical} and~\ref{eq:damping} are consistent as $m\rightarrow\infty$, since $(\pi \beta)^{1/m}\simeq 1$ for large values of $m$.

For large $n$ the critical curve constrains the maximum value of $V_1$ before the onset of parametric resonance in species A, as determined by \refeq{eq:damping}, where there is an implicit dependence of $m$ on $n = \OmH / \OmL$.
At small $n$, the stability regions between the $n-1$ tongues are wide and \refeq{eq:damping} is not valid.
However, it is possible to avoid parametric resonances by careful choice of trap parameters on a case-by-case basis, where more general rules cannot be established.
We note that trapping with two frequencies is most suitable for combinations of species with widely different $Q/M$, for which $n$ will be large.

\subsection{Enhancing the weakest spring constant}

We now discuss the parameter range for which the use of a second frequency increases the spring constant for the heavier species, without significantly affecting the lighter ions.
Comparison of the ratio of the spring constants for  single-frequency operation of the Paul trap (\refeq{eq:KBKA}) with that given in \refeq{eq:KBKALoHi} shows that the two-frequency scheme enhances the ratio $\kappa_B/\kappa_A$ by a factor of
\begin{equation}\label{eta}
\eta_{\mathrm{2rf}}= \left(\frac{V_1}{\Omega}\cdot\frac{\Omega_n }{V_n}\right)^2 = \left(\frac{n V_1}{V_n}\right)^2.
\end{equation}
The threshold voltage for parametric excitation of species A, given by \refeq{eq:damping}, can be rewritten as
\begin{equation}\label{V1Vn}
\frac{V_1}{V_n} = \frac{\rho}{2}\left(\frac{V_n}{2 V_A}\right)\left(\pi\beta\right)^{1/m} \equiv\zeta_0\left(\pi\beta\right)^{1/m}
\end{equation}
where $m$ is the order number and $\beta = \Gamma_A / \Omega$ describes the damping for motion driven at angular frequency $\Omega$, and we assume $\pi\beta\ll 1$; the function $\left(\pi \beta \right)^{1/m} \to 1$ as $m$ increases.
The quantity $V_n/2 V_A = q_A$ is the $q$-parameter in the Mathieu equation for species A for which we take the typical value of $q_A \simeq 0.4$ (and similarly for $q_B = V_1/2 V_B$) so that $\zeta_0 = \rho q_A/2 \simeq 0.1$ under optimal conditions.
Substituting \refeq{V1Vn} into \refeq{eta} gives
\begin{equation}\label{etage1}
\eta_{\mathrm{2rf}}=\left(\zeta_0 n\right)^2\left(\pi\beta\right)^{2/m}\simeq \left(\frac{n}{10}\right)^2\left(\pi\beta\right)^{2/m}.
\end{equation}
Hence, we require $n\gg 10$ for the two-frequency scheme to  enhance the trapping of two species, i.e.\ $\eta_{\mathrm{2rf}}>1$.
As an example consider typical values of $n=100$, $m=28$, and $\beta = 10^{-5}$, which give
$\left(\pi\beta\right)^{2/m}=0.48$ and $\eta_{\mathrm{2rf}}\simeq 48$.
Although the lower limit for $n$ could be estimated more accurately (using the relationship between $m$ and $n$ in \refeq{eq:mest}), it is more instructive to examine the two-frequency scheme from the viewpoint of balancing the spring constants.

\subsection{Range of applicability: balancing the spring constants}
\label{sec:range}

\begin{figure}[t]
\centering
\includegraphics[width=\columnwidth]{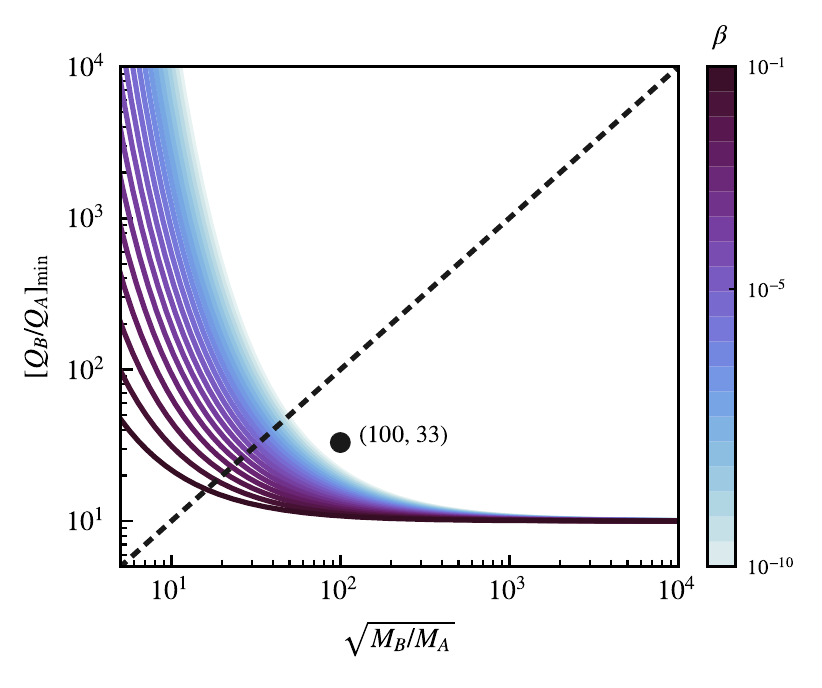}
\caption{The curves show the minimum charge ratio $Q_B/Q_A$ required to trap two ions of mass $M_A$, $M_B$  with the same spring constants, i.e.\ $\left[ \kappa_B/\kappa_A\right]_\mathrm{2rf}=1$; these are given by \refeq{QAQBbound} for $10^{-10} < \beta < 10^{-1}$. The value of damping becomes unimportant at large mass ratios  and  the minimum charge ratio tends to the asymptotic value given in \refeq{QAQBinfty}.
The black dashed line, $Q_B/Q_A = \sqrt{M_B/M_A}$, corresponds to \refeq{eq:KBKA} with the condition $\left[ \kappa_B/\kappa_A\right]_\mathrm{1rf}=1$. Using a  second frequency is most useful below this line. The point $(\sqrt{M_B/M_A}, Q_B/Q_A) = (100,33)$ corresponds to the numerical simulation in \refsec{sec:md}.}
\label{fig:chargeRatioConstraint}
\end{figure}
In this section, we find operating parameters for which the use of a second frequency ensures both species have similar spring constants in the radial direction of a linear Paul trap (see \ref{sec:twopaul}).
The same considerations apply for a quadrupole guide.
For this, the two-frequency scheme can be useful only when the ratio of the spring constants for single-frequency operation of the Paul trap (given in \refeq{eq:KBKA}) is less than unity, i.e.\ weaker trapping of the heavier species.
From \refeq{eq:KBKALoHi}, we find that balancing the spring constants, $\left[ \kappa_B/\kappa_A\right]_\mathrm{2rf}=1$, requires
\begin{equation}\label{QB2MB}
\left(\frac{nV_1}{V_n}\right)^2 \cdot \frac{Q_B^2/M_B}{Q_A^2/M_A} = 1.
\end{equation}
The maximum value of the voltage $V_1$ applied to the electrodes at the lower frequency dictates a minimum value of $Q_B^2/M_B$ for which the spring constants can be balanced, which can be found by substituting \refeq{V1Vn} in \refeq{QB2MB}.
We combine this bound on the ratio of $Q^2/M$ with the approximate relation
\begin{equation}
\label{eq:xi}
\frac{q_B}{q_A} = \frac{V_1/2V_B}{V_n/2V_A}\simeq 1,
\end{equation}
which assumes optimal trapping for both species, and
\begin{equation}\label{V1Vn2}
  \frac{V_B}{V_A} = \frac{M_B/Q_B}{n^2M_A/Q_A},
\end{equation}
to derive separate conditions for the charge and mass of the two species: $n \simeq \sqrt{M_B /M_A}$ and $Q_BV_1 \simeq Q_A V_n$.
Hence from \refeq{V1Vn} we find the lower bound of the charge ratio for which the spring constants can be made equal is given by
\begin{equation}\label{QAQBbound}
\frac{Q_B}{Q_A} \geq \left[\frac{Q_B}{Q_A}\right]_\infty \left(\pi\beta\right)^{-1/m}
\end{equation}
where the value as $m\rightarrow\infty$ is
\begin{equation}\label{QAQBinfty}
\left[\frac{Q_B}{Q_A}\right]_\infty = \frac{1}{\zeta_0}= \frac{2}{\rho q_A} \simeq 10.
\end{equation}
The order number is  $m\simeq q_A\sqrt{M_B/2M_A}$, from \refeq{eq:mest}.
The values of $[Q_B/Q_A]_{\mathrm{min}}$ are shown as a function of $n\simeq\sqrt{M_B/M_A}$ in \reffig{fig:chargeRatioConstraint} for various values of the damping parameter $\beta$.
Two-frequency trapping significantly improves the ratio of the spring constants for charge and mass values below the diagonal dashed line.
The lower bound does not depend on the masses $M_A$, $M_B$; only the ratio of the charges determines whether we can balance the spring constants using two frequencies at this universal limit.

\section{Experimental measurements}
\label{sec:exp}

\begin{figure}[t]
\centering
\includegraphics[width=0.8\columnwidth]{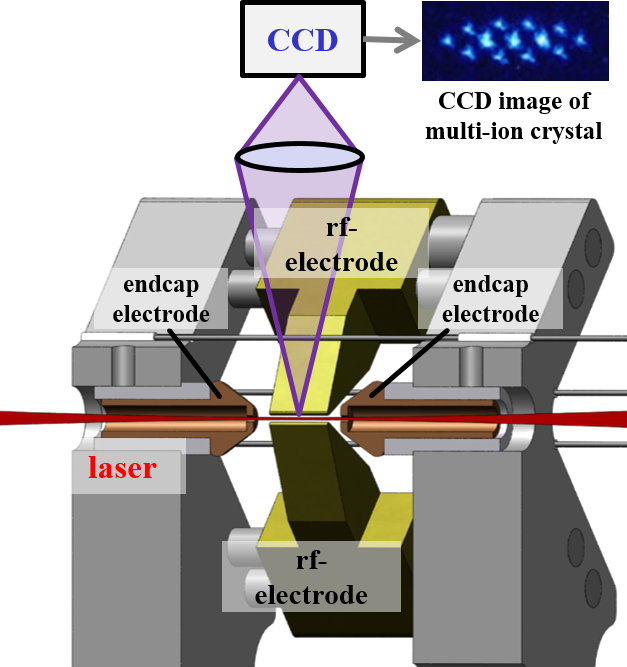}
\caption{The linear ion trap apparatus used for measuring the parametric resonances. The figure shows a cross-section through the trapping structure to expose the details of the rf and endcap electrodes. The trapped ions are imaged using a microscope lens and a EMCCD camera at an imaging direction perpendicular to the axis of the trap. The inset shows a typical image of a three-dimensional ion crystal.}
\label{fig:Setup1}
\end{figure}

\begin{figure}[ht]
\centering
\includegraphics[width=0.9\columnwidth]{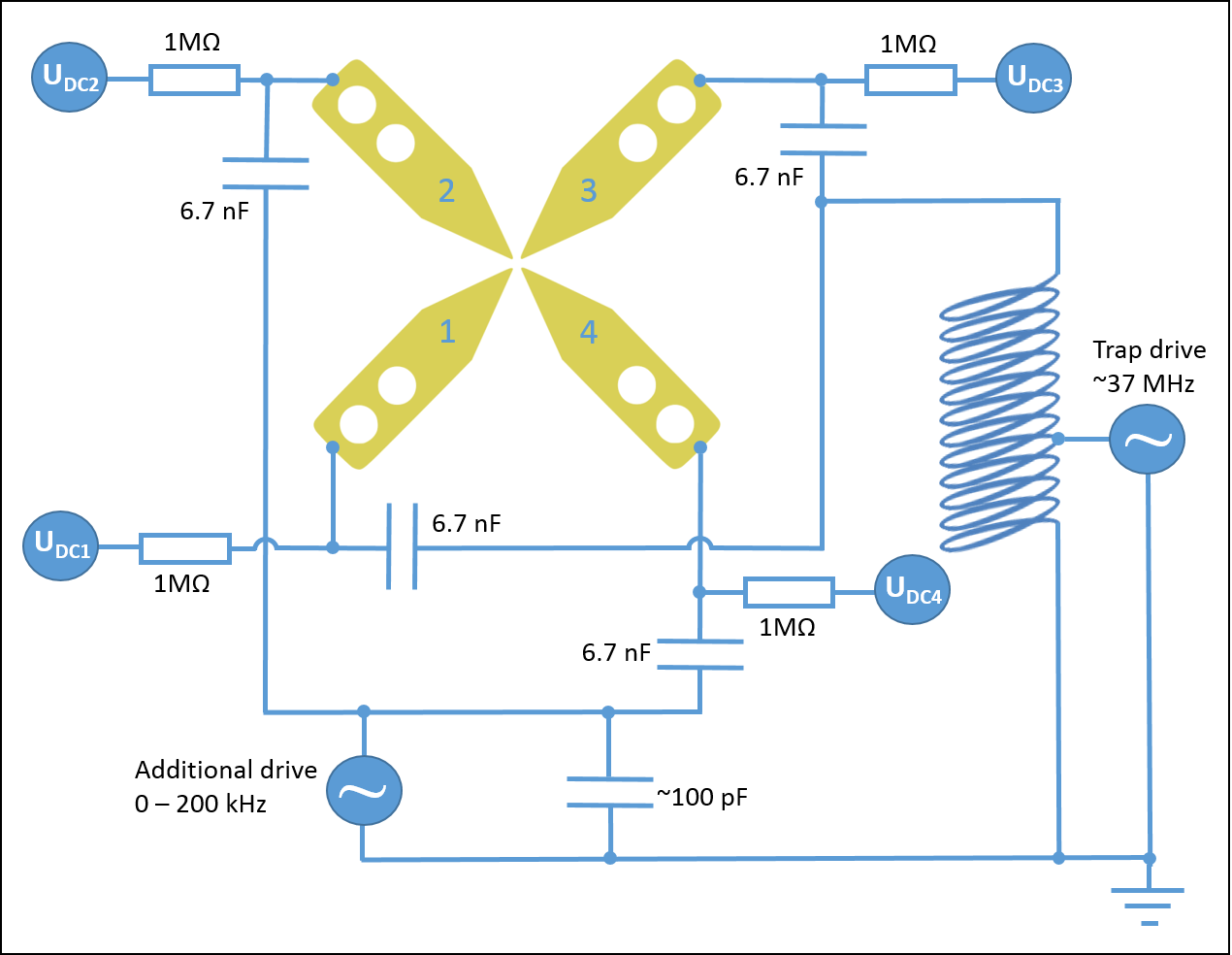}
\caption{Electrical circuit for two frequency operation of the ion trap. Two of the rf electrodes are driven at a high frequency whereas the second electrode pair is used for the low frequency drive. The dc and rf drives are decoupled using a series of capacitors.}
\label{fig:Setup2}
\end{figure}

We have carried out experimental measurements with \ca ions confined in a linear Paul trap to test the effect of parametric resonances on the stability of the atomic ions.
Our Paul trap has a standard configuration of four parallel blade-shaped electrodes~\cite{gardner_precision_2014} creating the quadrupole field operating with a single frequency $\OmH$.
Figure~\ref{fig:Setup1} shows the details of the apparatus.
The rf electrodes are 4\,mm long with an ion-electrode separation of 0.46\,mm.
Two endcap electrodes separated by 6\,mm provide the axial confinement and are symmetrically placed around the trap centre.
The electrodes have a hole in their centre for optical access.
We drive the trap asymmetrically, by applying the rf voltage to two opposing electrodes while connecting the other two electrodes to rf ground (see \reffig{fig:Setup2}).
Using the trap electrodes as a capacitance of a resonance circuit, we supply a low-voltage rf drive to a tap of a transformer coil.
To allow the application of dc voltages to the rf electrodes, for compensating static electric fields, the rf drive goes through a capacitor thus decoupling the rf drive circuit from the dc voltages.
The compensation voltage goes through a 1\,M$\Omega$ resistor which decouples the dc supply from the rf circuit without deteriorating its Q-factor (Q = 46.6).

The radial secular oscillation frequency of the \ca ions is $\omega_A/2\pi=\SI{1.35}{MHz}$ corresponding to $q=0.1$.
The axial secular frequency is \SI{0.483}{MHz} corresponding to $a=\num{3.2e-4}$ for a static voltage of \SI{600}{V} on the end-cap electrodes.
The amplitude of the applied voltage at $\OmH/2\pi =\SI{37.49}{MHz}$ deduced from the value of $q$ and an ion-electrode distance of \SI{0.46}{mm} is $V_n = \SI{250}{V}$.
Two-frequency trapping is achieved by applying a low frequency, low voltage drive to the rf ground electrodes. Using a 100\,pF capacitor across the low frequency drive provides a low impedance ground connection for the high frequency trap drive.
\begin{figure}[ht]
\centering
\includegraphics[width=\columnwidth]{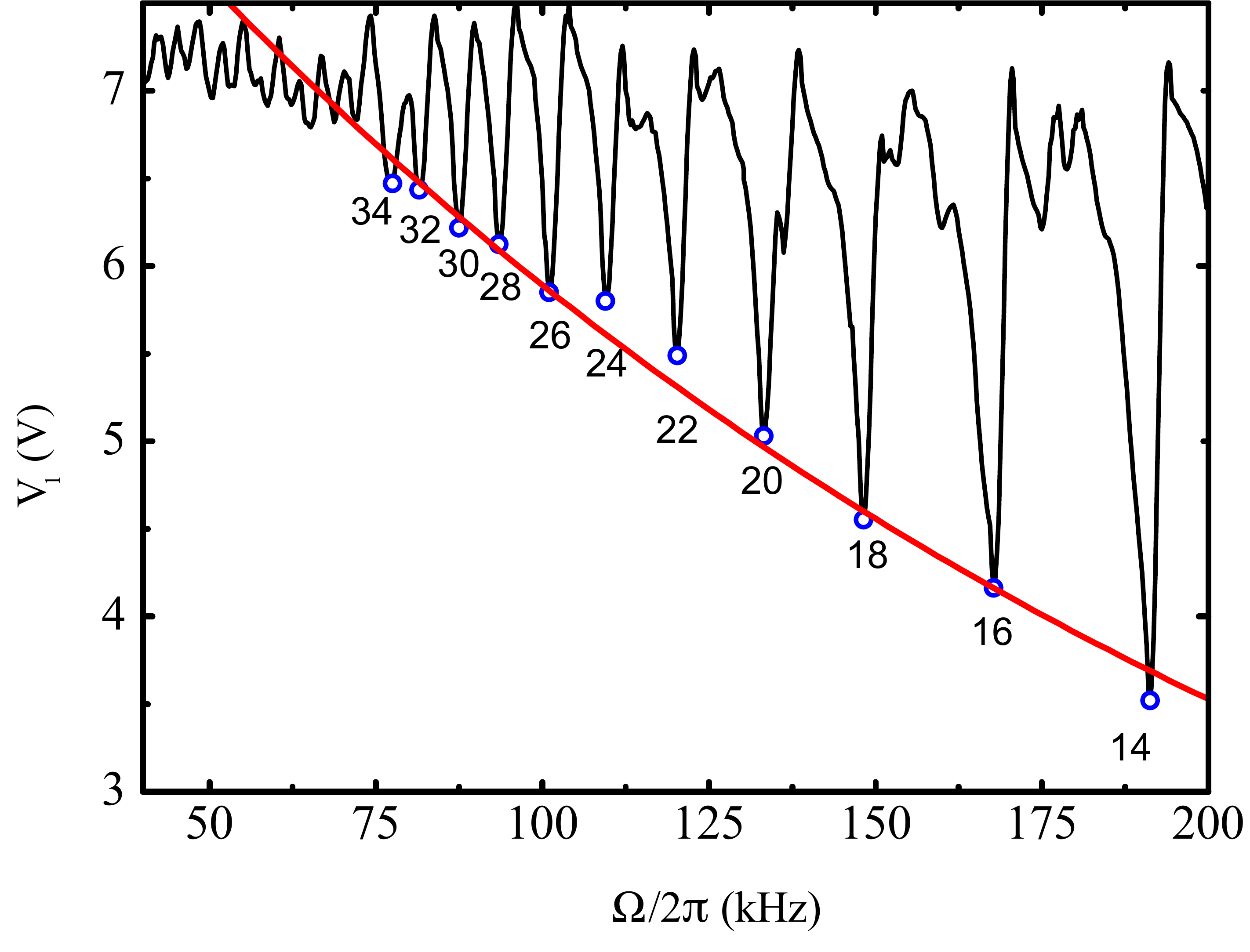}
\caption{Measured values of the threshold voltage versus the excitation frequency.
Twelve distinct resonances are labeled by their respective order number $m = 14,16,18,20,\dots,34$.
The higher order resonances, at lower frequencies, fade into the background noise.
The dots are the threshold voltage for each parametric resonance and the red line is \refeq{eq:damping} with best-fit parameters: $V_1=\SI{9.3}{V}$ for $\beta=10^{-6}$ however the maximum recorded threshold is 7.7\,V as $\Omega/2\pi\to 0$. The damping must be interpreted cautiously since only even orders are observed which is not predicted by the theory in the text.
}
\label{fig:exptdata}
\end{figure}

Our experiments started with loading a single \ca ion into the ion trap.
The measurements consisted of imaging the ion onto an EMCCD camera by capturing the fluorescent photons emitted when we excited the resonance transition with laser radiation at a wavelength of \SI{397}{nm}.
Throughout the measurements the ion was laser cooled close to the Doppler temperature when there was no parametric excitation.
We applied the lower frequency field and increased its amplitude until the Gaussian width of the fluorescent image of the ion increased to five pixels on the camera due to the ion's motion; a single pixel corresponded to a displacement of the ion by \SI{1.32}{\mu m}; we chose a cut-off width of five pixels to be well above the fluctuations of the ion's position.
This way we were able to take measurements without expelling the ion from the trap each time.
For each amplitude, we took two images and averaged their fitted widths along the $y$ radial direction.

The maximum voltage $\VL$ corresponding to the cut-off width is shown in \reffig{fig:exptdata} for the range $\OmL/2\pi = 40-200$ \,kHz.
There are twelve distinct resonances in this range, that correspond to even orders $m =14,16,\dots,34$.
The threshold for parametric excitation induced by the applied voltage $\VL$ increases as $\OmL$ decreases.
We fit the peaks of the resonances to determine values of $V_1$ and $\beta$.
The theoretical limiting voltage as $m\to\infty$ is $0.27 q_A V_n = \SI{7}{V}$ for $q_A = 0.1$ and $V_n = \SI{250}{V}$.
This is close to the measured maximum threshold \SI{7.7}{V}, but less than the best fit value of $V_1=\SI{9.3}{V}$.
We have not made any corrections for the effect of the finite value of $a$ arising from the dc voltage applied to the end-cap electrodes.
Moreover, even-order resonances were stronger than odd ones which is not expected from the theory for motion along one direction; fitting to even-order resonances only, overestimates the threshold voltage.
Similar odd-even staggering was seen in the measurement of ~\cite{razvi_fractional_1998}, but not in~\cite{zhao_parametric_2002,collings_observation_2000}.
Nevertheless, these measurements show that a voltage sufficient to confine the heavy ions can be applied in this $\OmL$ range for $\VL < \SI{7}{V}$.

Figure~\ref{fig:exptdata} also shows that it possible to avoid parametric resonances by a suitable choice of $\OmL$, although the presence of species B may shift the secular oscillation frequencies of species A.
For higher values of $m$, the resonances are so weak that their influence becomes comparable to the non-resonant effect of an additional dc voltage.
Thus, atomic ions are confined stably in a Paul trap with another quadrupole field oscillating at a lower frequency.

\section{Molecular dynamics simulations}
\label{sec:md}

We have carried out numerical simulations to confirm the stability of multiple ions in the two-frequency trap, in all three dimensions. These simulations use a time-varying electric field of the form
\begin{align}
\label{eq:MDfield}
\mathbf E(\mathbf x_j; t) &= \sum_{\kappa=1,n} \left(\frac{V_\mathrm{\kappa}}{r_0^2} \cos{\left(\Omega_\kappa t\right)} \left(y \hat{e}_y - x \hat{e}_x\right)\right) \\\nonumber
		&+ \frac{V_0}{z_0^2} \left(x \hat{e}_x + y \hat{e}_y - 2 z \hat{e}_z\right).
\end{align}
The oscillating voltages $V_k$ provide confinement in the $x$ and $y$ directions, while the static voltage $V_0$ confines ions in the $z$ direction.
The trajectories of $N$ interacting ions in the electric field of a Paul trap obey the equations of motion
\begin{equation}
M_j \ddot{\mathbf{x}}_j = \mathbf E(\mathbf x_j; t)Q_j + \sum_{\substack{i,j=1\\i\neq j}}^N\frac{1}{8\pi\epsilon_0}\frac{Q_iQ_j}{(\mathbf x_j-\mathbf x_i)^2}
\end{equation}
for ion $j$, where $E(\mathbf x_j; t)$ is the electric field of \refeq{eq:MDfield} arising from the two-frequency potential $V(t)=V_0+V_1\cos{\OmL t}+V_n\cos{\OmH t}$, and $\epsilon_0$ is the permittivity of vacuum.

We numerically integrated the equations of motion using (py)LIon~\cite{bentine_molecular_2014}, a collection of software tools we developed to investigate electrodynamic trapping of multiple species in an ion trap.
These tools provide a wrapper around LAMMPS~\cite{plimpton_fast_1995}~\footnote{LAMMPS is a classical molecular dynamics code developed by Sandia Labs and distributed as open-source software.}, exposing only a subset of its capabilities relevant to the simulation of ion trap dynamics.
The effect of collisions with a buffer gas is implemented by coupling the ions to a Langevin bath, which provides both a stochastic and viscous damping force that causes thermalisation of the ensemble to a specified temperature. A similar approach was previously used to simulate laser cooling in a single-frequency trap~\cite{zhang_molecular-dynamics_2007}.

\begin{figure}[t]
\centering
\includegraphics[width=\columnwidth]{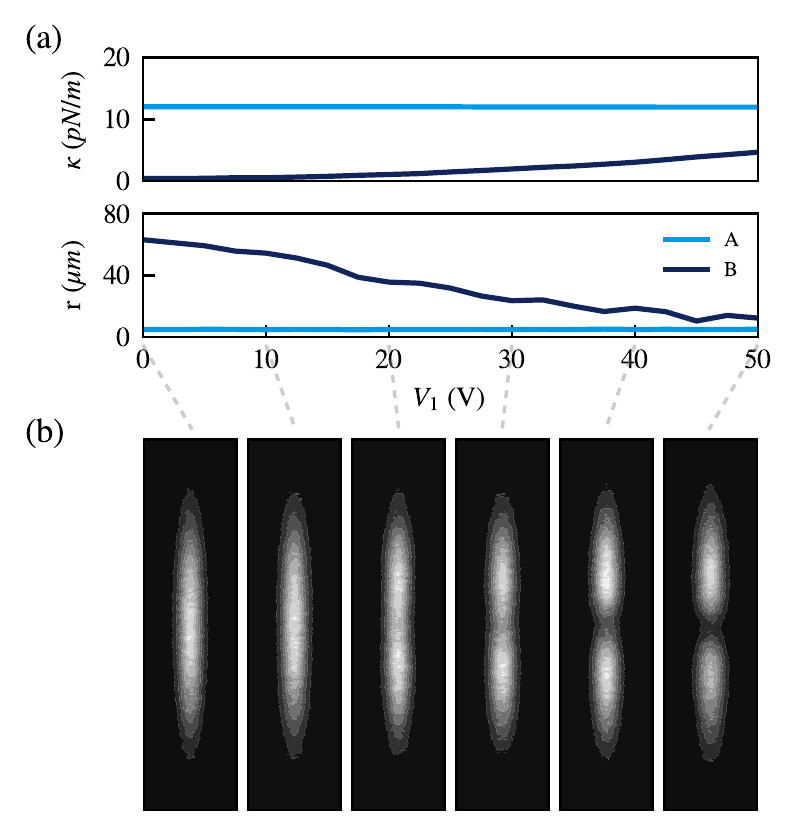}
\caption{
Molecular dynamics simulations of a cloud of 20 atomic ions and one massive ion in a linear Paul trap, for increasing values of the low-frequency drive voltage $V_1$.
This system's parameters are marked on \reffig{fig:chargeRatioConstraint}.
(a, top) the spring constants,  (a, bottom) the mean cloud radii of the two species, and (b) the density distribution of the fluorescing light species that develops a gap as the more massive ion is pulled inwards.
The values of $\kappa_A$ and $\langle r_A\rangle$ are approximately constant whereas $\kappa_B$ increases causing $\langle r_B\rangle$ to decrease.
}
\label{fig:md}
\end{figure}

For these simulations we used a linear Paul trap of length 2\,mm, radius 1.75\,mm and geometric factor $k = 0.325$ as defined in~\cite{berkeland_minimization_1998}.
The main driving frequency was $\Omega_n/2\pi = \SI{10.03}{MHz}$ and voltage $V_n = \SI{2.76}{kV}$ with an endcap voltage $V_0 = 2$\,V.
The lower driving frequency was $\Omega/2\pi = \SI{98}{kHz}$ with voltages up to 50\,V as shown in \reffig{fig:md}.
Note that $n=\Omega_n/\Omega\simeq 102$ is not an integer here.
With the trap configured as above, $q_A = (Q_A/M_A)\cdot 2\Vppm / \OmH^2 \simeq 0.32$ while $q_B \simeq 0.19$ for the maximum value of $V_1$.
The results of \reffig{fig:md} are for a system of 20 \ba ions and a single B ion with $M_B/M_A \simeq 10^4$ and $Q_B/Q_A = 33$, as marked on Fig.~\ref{fig:chargeRatioConstraint}.
The temperature of the Langevin bath that both species are in contact with is 10\,K, typical of buffer gas cooling, so the ions in this ensemble do not form a Coulomb crystal, and instead remain as thermal clouds.
The \ba ions are coupled to the bath with a velocity decay constant of \SI{100}{\micro\s}.

This specific example shows that applying two frequencies is useful for the particular case of \ba co-trapped with a much heavier ion of species B.
The large difference in $Q^2/M$ means that the species have  very dissimilar spring constants in an ion trap operated at a single frequency. Thus separation of the clouds occurs with weakly confined heavy ions pushed away from the trap centre by the Coulomb repulsion of the tightly bound light ions (see \reffig{fig:md}).
The maximum value of $V_1$ in \refeq{eq:damping} determines the confinement of species B through the parameter $q_B$.
Using \refeq{eq:KBKALoHi} we find a range of voltages $V_1$ where the two-frequency trap is stable but the ratio of the spring constants is dramatically different as shown in \reffig{fig:md}.
This increased overlap is possible only for certain values of ($Q_B$, $M_B$) since the condition that there is no parametric heating of species A by the lower frequency, limits the strength of confinement arising from the field produced by the applied voltage $V_1 \cos\left(\OmL t\right)$ as given by \refeq{QAQBbound}.
For a modest voltage of 50\,V the mean position of the heavy ion is very close to that of the lighter ones and at least an order of magnitude smaller than when $V_1 = 0$.

The simulated images in \reffig{fig:md} depict the position distribution of fluorescing \ba ions.
The presence of the heavy ion inside a cloud of lighter ones can be inferred by the dark hole that appears in the simulated fluoresence images when the second electric field is large; this occurs due to Coulomb repulsion of the light ions by the heavy one as it is pulled towards the centre of the trap.
We have obtained similar results for simulations with 200 \ba ions co-trapped with a single ion of species B.
Moreover, the ion of species B does not need to be in direct contact with the Langevin bath for these results to hold since it is being sympathetically cooled by the lighter ions at an increased efficiency compared to the single-frequency case.
Simulations with the damping term removed from species B show similar results to \reffig{fig:md}.

\section{Conclusions and outlook}
\label{sec:concl}

We have shown that two-frequency operation of a Paul trap or quadrupole guide can provide a significant advantage when trapping species with dissimilar charge to mass ratios.
In considering the stability of each species, we proposed an approach to finding suitable operating conditions within the multidimensional space.
We have verified the predicted stability of the atomic ions experimentally and carried out numeric simulations of the two-frequency, two-species system.
This previously unexploited mode of electrodynamic trapping, can be implemented on an existing apparatus and has general applicability.

Electrodynamic confinement has many uses ranging from mass spectrometry of small molecules to the control of particles of dimensions hundreds of micrometres.
We consider two broad categories of prospective applications: detection of heavy ions via their effect on fluorescing atomic ions, and reactions of cold molecular ions in cold (bio-)chemistry.

Detection of ions via their effect on fluorescing atomic ions being excited with laser light is used for dark ions such as molecules, or atomic ions with transitions at inconvenient  wavelengths.
Our results show how to extend this detection technique to massive ions with lower charge-to-mass ratios than atomic ions.
This can be used in mass spectrometry for the non-destructive detection of individual molecular ions at a low count rate.
Moreover, an intrinsic part of mass spectrometry is fragmentation, i.e., observing the breaking apart of a large biological complex, and the two-frequency scheme enables such investigation of single biomolecular complexes without ensemble averaging.
The efficient, non-destructive detection of biomolecular ions could be followed by controlled deposition of mass-selected particles on a surface for further analysis by other techniques.
It has also been suggested that a chain of atomic and molecular ions can act as a conveyor belt to carry the dark molecular ions into the focus of an x-ray laser for destructive measurements~\cite{kahra_molecular_2012}, as demonstrated in experiments with Mg$^+$ and MgH$^+$ which have similar charge-to-mass ratios.
Our method opens the way to working with much heavier biomolecular ions.

Paul traps are used to investigate chemical reactions of ions at low temperatures~\cite{bell_ultracold_2009}, and two-frequency operation allows much heavier species to be used without a large spatial separation.
This application is closely related to the use of two frequencies for the confinement of antimatter where the creation of overlapping clouds of positrons and anti-protons leads to the formation of anti-hydrogen atoms; work towards this is technically challenging goal is ongoing~\cite{leefer_investigation_2016}\footnote{
Sympathetic cooling of positrons by Be$^+$ ions has been demonstrated experimentally in a Penning trap~\cite{jelenkovic_sympathetically_2003}.}.

Sympathetic cooling of biomolecular ions by \ba in a Paul trap was demonstrated \cite{offenberg_translational_2008} and heavier ions can be used by applying  two frequencies.
The transfer of energy between ions is less efficient if the oscillation frequencies are dissimilar.
However, it is possible to arrange a system with coupling between axial modes of species A and radial modes of species B in a linear Paul trap~\cite{trypogeorgos_cotrapping_2016}.
The extension of the techniques developed for laser-cooled atomic ions to species with much higher $M/Q$ that are sympathetically cooled presents new ground for future investigation.
Our method allows for cotrapping atomic ions with much heavier charged particles, such as flakes of graphite~\cite{kane_levitated_2010-1} or DNA molecules in aqueous solutions~\cite{zhao_molecular_2008}.

\section*{Acknowledgements}
This work was supported by the EU H2020 Collaborative project QuProCS (Grant
Agreement 641277). DT and EB acknowledge support from the EPSRC. DT is thankful to the Bodossaki foundation and St. Peter's College. We also gratefully acknowledge support from EPSRC (EP/J003670/1).



\appendix

\section{The mathematical treatment of damping}
\label{sec:damp}

Equations of motion with damping proportional to velocity, or any second-order differential equation in which the first and second derivatives have constant coefficients, can be written in the general form $\ddot y + \Gamma \dot y + H(t) y =0$.
The substitution $ y =  \tilde y \exp( -\Gamma t/2)$ eliminates the term with the first derivative to give $\ddot{\tilde y} + \tilde H(t)  \tilde y =0$,
where $\tilde H(t) = H(t) - \Gamma^2/4$.
Second-order equations with linear damping have an associated equation without damping which is a Hill equation when $\tilde H(t)$ is a periodic function and can be written as the sum of its Fourier components.
An illustrative example is the damped simple harmonic oscillator with a parametric driving term
\begin{equation}\label{eq:Blumel4tilde}
\frac{d^2 y}{dt^2} + \tilde\omega_A^2\, y= \frac{\Fp}{M_A} \cos \left(\Omega t\right) y,
\end{equation}
where $\Fp$ is a constant and the secular oscillation frequency is modified to $\tilde\omega_A^2 = \omega_A^2 - \Gamma^2/4$.
This is the usual shift of the resonance frequency that occurs in damped simple harmonic motion.
The integrating factor $\exp( - \Gamma t/2)$ eliminates $\dot y$ for any value of the constant $\Gamma$ and is used in numerical calculations of the stability regions of the Mathieu equation with strong damping~\cite{nasse_influence_2001, hasegawa_dynamics_1995}.

To relate the behaviour of ions in a quadrupole electric field oscillating at two frequencies (including a dc term) to familiar properties of a single-frequency Paul trap we consider \refeq{eq:yA} with $V_1=0$ which reduces to a Mathieu equation with parameter
\begin{equation}\label{eq:qAVn}
\qt = \frac{V_n}{2V_A} = \frac{Q_A}{M_A} \cdot \frac{4 V^{\prime\prime}_n}{ (n\Omega)^2},
\end{equation}
where $V^{\prime\prime}_n = V_n/r_0^2$ is the curvature of the oscillating electric potential.
We drop the minus sign since the steady-state behaviour does not depend on the sign of $q$.
We take $\qt = 0.4$ as a realistic value for operation of a Paul trap since for higher values the ions are susceptible to heating and loss from nonlinear mixing and parametric excitation even for single frequency operation.

The quadrupole field proportional to $V_n \cos n\Omega t$ leads to a pseudopotential with a secular oscillation frequency $\omega_A = \qt n\Omega/\sqrt{8}$ so ions of species A undergo simple harmonic motion described by $\ddot y + \omega_A^2 y =0$ where $t$ is real time.
We introduce a voltage $V_1 \cos \Omega t$ that produces a driving term proportional to $\Fp = Q_A  V_1/r_0^2$ in \refeq{eq:Blumel4tilde}.
Using the rescaling $\Omega t = 2\tau$ as in \refeq{eq:yB} gives
\begin{equation}\label{eq:Blumel7}
\frac{d^2 y}{d\tau^2} +2\beta \frac{d y}{d\tau} + \left(a_{\mathrm{eff}} -2q_A^{\{\tau\}} \cos 2 \tau\right) y =0,
\end{equation}
where $\beta =\Gamma_A/\Om$ and
\begin{equation}\label{eq:aeff}
a_{\mathrm{eff}} = \left(\frac{2\omega_A}{\Omega} \right)^2 =  \frac 12 \left( \frac{n V_n}{2 V_A} \right)^2 = \frac 12 \left(n \qt \right)^2
 \end{equation}
is the effective value of a static potential equivalent to the pseudopotential (with $V_0=0$), and
\begin{equation}\label{eq:qtau}
-\qtau = \frac{V_1 n^2}{2V_A}
\end{equation}
is completely independent from $\qt$.
For $\qt =0.4$ and $n=100$ we find that $a_{\mathrm{eff}}=800$.
This is far greater than is usually considered in the theory of Paul traps but the asymptotic properties of the Mathieu equation are known in literature~\cite{broer_geometrical_1995,broer_large_2013}.
For $a, q\gg 1 $ the critical line between predominantly stable and unstable regions is $a =2q$, as shown in Fig.~\ref{fig:1}a).
This passes through the point $(q,a)=(400, 800)$ in this example.
Using $a_{\mathrm{eff}} = 2\vert \qtau\vert$, and Equations~\ref{eq:aeff},~\ref{eq:qtau}, we recover \refeq{eq:critical} with the same value $\rho=0.5$ as the simplified argument given previously.
Damping modifies $\aeff$ however this is not a significant effect for $\Gamma/(2\omA) <0.1$ and is neglected here.

\section{Theory of parametric excitation with linear damping}
\label{sec:damppe}

The parametric excitation of ions in a Paul trap was investigated experimentally and theoretically in~\cite{razvi_fractional_1998,zhao_parametric_2002}.
We extend their results to high order resonances using asymptotic properties of the instability tongues of the Mathieu equation.
The threshold voltage above which an applied field at $\Omega$ excites a resonance of order $m$ is
\begin{equation}\label{eq:Schuess16}
\left[ \frac{V_1}{r_0^2} \right]_{\mathrm{th}} = \frac{2 M \omega_A^2}{Q }C_m \left(\frac{\Gamma}{2\Omega}\right)^{1/m},
\end{equation}
with $M\equiv M_A$, $Q\equiv Q_A$ and $\Gamma \equiv \Gamma_A$ here.
This corresponds to \refeq{eq:damping} with $\rho = (2/ \mathrm{e})^2 \simeq 0.54$ for
\begin{equation}\label{eq:Cm}
C_m = \left(\frac{2}{e} \right)^{2}(\pi m)^{1/m}.
\end{equation}
This formula for $C_m$ is derived below by relating the damping to the width of tongues of instability.
Razvi \emph{et al.}~\cite{razvi_fractional_1998} estimated the coefficient to be $C_m\simeq 2$ from numerical calculations of the first few resonances.
Zhao \emph{et\,al.}~\cite{zhao_parametric_2002} find the values $\{C_1, C_2, \dots  C_6\} = \{2, 1.414, \dots 0.887 \}$ whereas \refeq{eq:Cm} gives $\{1.7, 1.36, \dots 0.887 \}$.
Thus our general formula is a good approximation for the width of resonances except for $m = 1$.
Importantly it gives an analytic expression for the high orders $m\gg 10$ relevant to our two-frequency scheme.

The following derivation uses notation similar to Zhao \emph{et\,al.} to highlight similarities and differences.
The damping constant $\beta =\Gamma/\Omega$ can be eliminated from \refeq{eq:Blumel7} by the substitution $y=\tilde y \exp(-\beta \tau )$ to give the Mathieu equation
\begin{equation}\label{eq:MathieuApp}
\frac{d^2 \tilde y}{d\tau^2} + \left(\tilde a_{\mathrm{eff}} -2 \qtau \cos 2 \tau\right) \tilde y =0,
\end{equation}
where $\tilde a_{\mathrm{eff}} = (a_{\mathrm{eff}} - \beta^2)$ hence
\begin{equation}
\tilde a_{\mathrm{eff}} = \left(\frac{2\omega_A}{\Omega} \right)^2 (1 -\gamma^2) = \left(\frac{2\tilde \omega_A}{\Omega} \right)^2,
\end{equation}
with $\gamma = \Gamma/(2\omega_A) \ll 1$ and $\tilde\omega_A =\omega_A\sqrt{1 -\gamma^2}$.
We neglect the slight frequency shift $(\omega_A - \tilde \omega_A)$ because $\gamma^2 \ll 1$.
Using \refeq{eq:mest} we find
\begin{equation}\label{eq:qepsilon}
\qtau = \frac{ 2 Q V_1 }{M r_0^2 \Omega^2} = \epsilon\left(\frac{2\omega_A}{\Omega} \right)^2 = \epsilon\, m^2
\end{equation}
is unaffected by damping, and we have defined
\begin{equation}\label{eq:epsilon}
\epsilon= \frac{Q V_1 }{2 M r_0^2 \omega_A^2}.
\end{equation}
Floquet's theorem states that equations with periodic coefficients have solutions of the form $\tilde{y}= e^{\mu\tau}u(\tau)$
where $u$ is a periodic function; in this case $u(\tau)=u(\tau +\pi)$ to match the period of $\cos 2 \tau $.
The solution of \refeq{eq:MathieuApp} is a linear combination of two such independent functions.
Hence the solutions in real time $t$ have the form
\begin{equation}
 y= C_1 e^{(\mu\Om - \Gamma)t/2}u_1(t) + C_2 e^{-(\mu\Om+\Gamma)t/2}u_2(t).
\end{equation}
Instability arises if $\mu > \Gamma/ \Om  = \gamma\,m$.
The stability of the solution can be determined from the relationship between the characteristic exponent $\mu$ and the width of the instability tongues.
Considering the shape of the tip of the tongues gives
\begin{equation}
\gamma = \frac{\mu}{m} = \frac{\aplus_m -\aminus_m}{4m^2},
\end{equation}
where $\aplus_m (q)$ and $\aminus_m (q)$ are the values at the upper and lower boundary of the $m$\,{\small th} tongue.
The width of the undamped resonance is $\Delta a_m = \aplus_m -\aminus_m = A_m q^m$.
From \refeq{eq:qepsilon} we find the excitation threshold as
\begin{equation}\label{eq:eth}
\left[\, \epsilon\, \right]_{\mathrm{th} } = \left(\frac{4 m^2}{A_m}\right)^{1/m}\frac{\gamma^{1/m} }{m^2}=C_m \gamma^{1/m}.
\end{equation}
This is equivalent to \refeq{eq:Schuess16}, with
\begin{equation}
C_m = \frac{1}{m^2} \left(\frac{4m^2}{A_m}\right)^{1/m}.
\end{equation}
Using the asymptotic formula for $A_m$ given in~\cite{noauthor_nist_nodate}
\begin{equation}\label{eq:Am}
\frac{A_m}{4m^2} = \frac{2} {2^{2m} [m!]^2},
\end{equation}
and Sterling's formula for $m!$
\begin{equation}
A_m = \frac{4}{\pi}\left(\frac{\mathrm{e}}{2m}\right)^{2m}
\end{equation}
leads to \refeq{eq:Cm}.

We can now use this to examine the validity of the approximations.
Resonances occur for $a_m=m^2$ hence their spacing is approximately $a_{m+1} -a_m \simeq 2m$ for large $m$.
Thus the condition that width of the tongues is small compared to their spacing is $A_m q^m \ll 2m$ which implies  $q \ll (\pi m /2)^{1/m}(2m/\mathrm{e} )^2$.
This is satisfied if  $2q \ll a =m^2$ which corresponds to the expectations that the approximation is valid well away from the critical line where the tongues of instability are narrow.
The formula for the width $\Delta a_m = \aplus_m -\aminus_m = A_m q^m$ is not restricted to small values of $q$; \refeq{eq:critical} is a limiting form of \refeq{eq:damping}.
For stronger damping, $\Gamma >0.1$, other approximations may be useful, or in such cases it is straightforward to carry out numerical calculations since the tongues of instability have rounded ends so that computing their boundary does not require an excessively fine computational grid.
In contrast the stability regions reaching into the region $2q >a$ are cusped even for strong damping but these are not relevant here.

\section{Two species in a linear Paul trap}
\label{sec:twopaul}

A linear Paul trap has four electrodes aligned parallel to the $z$-axis arranged on the corners of a square in the $xy$-plane, with adjacent electrodes having opposite polarity in a quadrupole configuration.
It is assumed that the trap is operated at the maximum voltage in a given apparatus to give the strongest confinement but this might not be optimal, e.g., for very large clouds of ions.
The oscillating electric field has no component of the electric field along $z$ and $E_y = \Vppm y $ then $E_x = - \Vppm x$ where $\Vppm =V/r_0^2$ is the curvature.
The Mathieu equations describing the motion in the $x$ and $y$ directions respectively $a_x = -a_y$. If $q=0$ then \refeq{eq:Mathieu} simplifies to simple harmonic motion $ y^{\prime\prime} = -a y$ at angular frequency $\sqrt{a}$ for $a>0$, and for $a<0$ the motion is unstable.
Thus the central axis of the four electrodes is a line of saddle points of the electrostatic potential energy of charged particles commonly referred to as the rf-null line.
There can be stable motion when an oscillating voltage is applied to these electrodes.
This can be shown using the pseudopotential approximation, valid for small  $a$ and $q$ in Mathieu equation, by substituting a trial solution of the form $ y = C \cos(\omega t)(1+D\cos(\Omega t))$, where $t$ is the real time.
Equating terms with the same time dependence leads to \refeq{eq:omegaaq}.
The pseudopotential has an effective depth of $qV_{osc}/8$; the Dehmelt approximation~\cite{march_quadrupole_2005}.
The discussion of radial confinement in the main text can be summarised by expressions for the potential energy of ions along the $y$-axis:
\begin{align}\label{eq:Ukappa}
U_A &= \left( \kappa_A(0) +\kappa_A(\OmH) \right) y^2/2, \\
U_B &= \left( \kappa_B(0) +\kappa_B(\OmL) +\kappa_B(\OmH) \right) y^2/2, \nonumber \\
U_B &\simeq \left( \kappa_B(0) +\kappa_B(\OmL) \right) y^2/2.
\end{align}
The last line follows because $\kappa_B(\OmL) \gg \kappa_B(\OmH)$ in two-frequency operation.

In the main text we considered that the dc terms are negligible: $\kappa_B(0) \simeq 0 \simeq \kappa_A(0)$.
In a linear Paul trap, however, there is a static radial field as a consequence of the axial confinement.
An axially symmetric voltage satisfying $\nabla\mathbf{E} =0$ has the form $ \Vz [z^2 -\harf(x^2+y^2)]/(2z_0^2)$ hence the axial confinement necessarily acts oppositely in the radial directions, e.g., $\kappa_{B}(0)\equiv \kappa_{B,y}(0) =-\kBz/2$.
The static spring constant $\kBz = Q_B\Vz/z_0^2$ depends only on the charge, and similarly $\kAz = Q_A\Vz/z_0^2$, which is very different to the dependence on $Q^2/M$ for an ac field.
The anti-trapping arising from the radial component of the static field is of particular concern for species B which is only weakly confined by a single-frequency ac field.
Consider a Paul trap with an aspect ratio $\lambda = \omA/\omega_{A,z}$; this ratio of radial to axial oscillation frequencies for species A gives the ratio of the length to radius of the elongated cigar-shaped cloud of A-ions in thermal equilibrium $\kappa_A(\OmH) =\lambda^2 \kAz$.
The requirement that $\kappa_B(\OmH) \ge \kBz/2$ gives a stability condition for species B in a single-frequency linear trap
\begin{equation}\label{eq:Bstability}
\frac{Q_B}{M_B}\ge \frac{1}{2\lambda^2}\cdot\frac{Q_A}{M_A},
\end{equation}
where we have used \refeq{eq:KBKA}.
For two species with $(Q_B/M_B)/(Q_A/M_A)= 3.3\cdot 10^{-3}$ this gives $\lambda \ge 12.3$.
Hence this large aspect ratio is required to reach the edge of the stability region, indicating the difficulty of confining two species of widely different $Q/M$ with a standard Paul trap.
Also this underestimates the problem since a cloud of atomic ions of species A collectively exert a stronger repelling force on an ion of species B.
The collective effect of many ions of species A should also be taken into account in two-frequency operation since it acts to drive species B out of the trap.
Another factor for a linear Paul trap is that there is the weaker confinement along the $z$-axis than radially $\kappa_{A,z} < \kappa_{A,y} \sim \kappa_{A,x}$ and therefore the criterion for species B to lie on the axis is $\kappa_{B,y} > \kappa_{A,z}$.
The heavier ion(s) can displace species A along the $z$-axis more easily than radially as shown in \reffig{fig:md}.
Thus, as stated in the main text, the criterion $\kappa_B \ge \kappa_A$ which considers only the radial direction(s) is a simplification.
Nevertheless, these calculations provide useful physical insight; the various competing effects can be studied using MD simulations.



\bibliographystyle{elsarticle-num}
\bibliography{twofreq_mass_spec_no_url,tobePublished}

\begin{thebibliography}{10}
\expandafter\ifx\csname url\endcsname\relax
  \def\url#1{\texttt{#1}}\fi
\expandafter\ifx\csname urlprefix\endcsname\relax\def\urlprefix{URL }\fi
\expandafter\ifx\csname href\endcsname\relax
  \def\href#1#2{#2} \def\path#1{#1}\fi

\bibitem{paul_electromagnetic_1990-1}
W.~Paul, Electromagnetic traps for charged and neutral particles, Rev. Mod.
  Phys. 62~(3) (1990) 531--540.
\newblock \href {http://dx.doi.org/10.1103/RevModPhys.62.531}
  {\path{doi:10.1103/RevModPhys.62.531}}.

\bibitem{paul_electromagnetic_1990}
W.~Paul, Electromagnetic {Traps} for {Charged} and {Neutral} {Particles}
  ({Nobel} {Lecture}), Angewandte Chemie International Edition in English
  29~(7) (1990) 739--748.
\newblock \href {http://dx.doi.org/10.1002/anie.199007391}
  {\path{doi:10.1002/anie.199007391}}.

\bibitem{earnshaw_nature_1848}
S.~Earnshaw, On the {Nature} of the {Molecular} {Forces} which {Regulate} the
  {Constitution} of the {Luminiferous} {Ether}, Transactions of the Cambridge
  Philosophical Society 7 (1848) 97.

\bibitem{hoffrogge_planar_2011}
J.~Hoffrogge, P.~Hommelhoff, Planar microwave structures for electron guiding,
  New J. Phys. 13~(9) (2011) 095012.
\newblock \href {http://dx.doi.org/10.1088/1367-2630/13/9/095012}
  {\path{doi:10.1088/1367-2630/13/9/095012}}.

\bibitem{nasse_influence_2001}
M.~Nasse, C.~Foot, Influence of background pressure on the stability region of
  a {Paul} trap, European Journal of Physics 22 (2001) 563--573.
\newblock \href {http://dx.doi.org/10.1088/0143-0807/22/6/301}
  {\path{doi:10.1088/0143-0807/22/6/301}}.

\bibitem{winter_simple_1991}
H.~Winter, H.~W. Ortjohann, Simple demonstration of storing macroscopic
  particles in a ‘‘{Paul} trap’’, American Journal of Physics 59~(9)
  (1991) 807--813.
\newblock \href {http://dx.doi.org/10.1119/1.16830}
  {\path{doi:10.1119/1.16830}}.

\bibitem{huang_gas-phase_2010}
T.-Y. Huang, S.~A. McLuckey, Gas-{Phase} {Chemistry} of {Multiply} {Charged}
  {Bioions} in {Analytical} {Mass} {Spectrometry}, Annu Rev Anal Chem (Palo
  Alto Calif) 3 (2010) 365--385.
\newblock \href {http://dx.doi.org/10.1146/annurev.anchem.111808.073725}
  {\path{doi:10.1146/annurev.anchem.111808.073725}}.

\bibitem{schiller_molecular_2003}
S.~Schiller, C.~Lammerzahl, Molecular dynamics simulation of sympathetic
  crystallization of molecular ions, Phys. Rev. A 68~(5) (2003) 053406.
\newblock \href {http://dx.doi.org/10.1103/PhysRevA.68.053406}
  {\path{doi:10.1103/PhysRevA.68.053406}}.

\bibitem{hilton_two_2012}
G.~R. Hilton, J.~L.~P. Benesch, Two decades of studying non-covalent
  biomolecular assemblies by means of electrospray ionization mass
  spectrometry, J. R. Soc. Interface 9~(70) (2012) 801--816.
\newblock \href {http://dx.doi.org/10.1098/rsif.2011.0823}
  {\path{doi:10.1098/rsif.2011.0823}}.

\bibitem{ostendorf_sympathetic_2006}
A.~Ostendorf, C.~B. Zhang, M.~A. Wilson, D.~Offenberg, B.~Roth, S.~Schiller,
  Sympathetic {Cooling} of {Complex} {Molecular} {Ions} to {Millikelvin}
  {Temperatures}, Phys. Rev. Lett. 97~(24) (2006) 243005.
\newblock \href {http://dx.doi.org/10.1103/PhysRevLett.97.243005}
  {\path{doi:10.1103/PhysRevLett.97.243005}}.

\bibitem{zhang_molecular-dynamics_2007}
C.~B. Zhang, D.~Offenberg, B.~Roth, M.~A. Wilson, S.~Schiller,
  Molecular-dynamics simulations of cold single-species and multispecies ion
  ensembles in a linear {Paul} trap, Phys. Rev. A 76~(1) (2007) 012719.
\newblock \href {http://dx.doi.org/10.1103/PhysRevA.76.012719}
  {\path{doi:10.1103/PhysRevA.76.012719}}.

\bibitem{offenberg_translational_2008}
D.~Offenberg, C.~B. Zhang, C.~Wellers, B.~Roth, S.~Schiller, Translational
  cooling and storage of protonated proteins in an ion trap at subkelvin
  temperatures, Phys. Rev. A 78~(6) (2008) 061401.
\newblock \href {http://dx.doi.org/10.1103/PhysRevA.78.061401}
  {\path{doi:10.1103/PhysRevA.78.061401}}.

\bibitem{dehmelt_economic_1995}
H.~Dehmelt, Economic synthesis and precision spectroscopy of anti-molecular
  hydrogen ions in {Paul} trap, Phys. Scr. 1995~(T59) (1995) 423.
\newblock \href {http://dx.doi.org/10.1088/0031-8949/1995/T59/060}
  {\path{doi:10.1088/0031-8949/1995/T59/060}}.

\bibitem{leefer_investigation_2016}
N.~Leefer, K.~Krimmel, W.~Bertsche, D.~Budker, J.~Fajans, R.~Folman,
  H.~Haeffner, F.~Schmidt-Kaler, Investigation of two-frequency {Paul} traps
  for antihydrogen production, arXiv:1603.09444 [hep-ex, physics:physics]ArXiv:
  1603.09444.

\bibitem{trypogeorgos_cotrapping_2016}
D.~Trypogeorgos, C.~J. Foot, Cotrapping different species in ion traps using
  multiple radio frequencies, Phys. Rev. A 94~(2) (2016) 023609.
\newblock \href {http://dx.doi.org/10.1103/PhysRevA.94.023609}
  {\path{doi:10.1103/PhysRevA.94.023609}}.

\bibitem{hasegawa_dynamics_1995}
T.~Hasegawa, K.~Uehara, Dynamics of a single particle in a {Paul} trap in the
  presence of the damping force, Applied Physics B Laser and Optics 61 (1995)
  159--163.
\newblock \href {http://dx.doi.org/10.1007/BF01090937}
  {\path{doi:10.1007/BF01090937}}.

\bibitem{pedersen_stability_1980}
P.~Pedersen, Stability of the solutions to {Mathieu}-{Hill} equations with
  damping, Ing. arch 49~(1) (1980) 15--29.
\newblock \href {http://dx.doi.org/10.1007/BF00536595}
  {\path{doi:10.1007/BF00536595}}.

\bibitem{berkeland_minimization_1998}
D.~J. Berkeland, J.~D. Miller, J.~C. Bergquist, W.~M. Itano, D.~J. Wineland,
  Minimization of ion micromotion in a {Paul} trap, Journal of Applied Physics
  83~(10) (1998) 5025--5033.
\newblock \href {http://dx.doi.org/10.1063/1.367318}
  {\path{doi:10.1063/1.367318}}.

\bibitem{wineland_experimental_1998}
D.~Wineland, C.~Monroe, W.~Itano, D.~Leibfried, B.~King, D.~Meekhof,
  Experimental issues in coherent quantum-state manipulation of trapped atomic
  ions, Journal of Research of the National Institute of Standards and
  Technology 103~(3) (1998) 259.
\newblock \href {http://dx.doi.org/10.6028/jres.103.019}
  {\path{doi:10.6028/jres.103.019}}.

\bibitem{ghosh_ion_1996}
P.~K. Ghosh, Ion {Traps}, Oxford University Press, Oxford : New York, 1996.

\bibitem{jordan_nonlinear_2007}
D.~Jordan, P.~Smith, Nonlinear {Ordinary} {Differential} {Equations}: {An}
  {Introduction} for {Scientists} and {Engineers}, 4th Edition, OUP Oxford,
  2007.

\bibitem{floquet_sur_1883}
G.~Floquet, Sur les equations differentielles lineaires a coefficients
  periodiques, Annales scientifiques de l'Ecole Normale Superieure 12 (1883)
  47--88.

\bibitem{konenkov_matrix_2002}
N.~V. Konenkov, M.~Sudakov, D.~J. Douglas, Matrix methods for the calculation
  of stability diagrams in quadrupole mass spectrometry, J Am Soc Mass Spectrom
  13~(6) (2002) 597--613.
\newblock \href {http://dx.doi.org/10.1016/S1044-0305(02)00365-3}
  {\path{doi:10.1016/S1044-0305(02)00365-3}}.

\bibitem{arnold_mathematical_1989}
V.~I. Arnol'd, Mathematical {Methods} of {Classical} {Mechanics}, Springer,
  1989.

\bibitem{weinstein_asymptotic_1987}
M.~I. Weinstein, J.~B. Keller, Asymptotic {Behavior} of {Stability} {Regions}
  for {Hill}'s {Equation}, SIAM Journal on Applied Mathematics 47~(5) (1987)
  941--958, articleType: research-article / Full publication date: Oct., 1987 /
  Copyright 1987 Society for Industrial and Applied Mathematics.
\newblock \href {http://dx.doi.org/10.2307/2101700}
  {\path{doi:10.2307/2101700}}.

\bibitem{simakhina_computing_2005}
S.~V. Simakhina, C.~Tier, Computing the stability regions of {Hill}'s equation,
  Applied Mathematics and Computation 162~(2) (2005) 639--660.
\newblock \href {http://dx.doi.org/10.1016/j.amc.2004.01.002}
  {\path{doi:10.1016/j.amc.2004.01.002}}.

\bibitem{roncaratti_whittakerhill_2010}
L.~F. Roncaratti, V.~Aquilanti, Whittaker–{Hill} equation, {Ince}
  polynomials, and molecular torsional modes, International Journal of Quantum
  Chemistry 110~(3) (2010) 716--730.
\newblock \href {http://dx.doi.org/10.1002/qua.22255}
  {\path{doi:10.1002/qua.22255}}.

\bibitem{broer_large_2013}
H.~Broer, M.~Levi, C.~Simo, Large scale radial stability density of {Hill}'s
  equation, Nonlinearity 26~(2) (2013) 565.
\newblock \href {http://dx.doi.org/10.1088/0951-7715/26/2/565}
  {\path{doi:10.1088/0951-7715/26/2/565}}.

\bibitem{landau_mechanics:_1976}
L.~D. Landau, E.~M. Lifshitz, Mechanics: {Volume} 1, 3rd Edition,
  Butterworth-Heinemann, Amsterdam u.a, 1976.

\bibitem{razvi_fractional_1998}
M.~A.~N. Razvi, X.~Z. Chu, R.~Alheit, G.~Werth, R.~Blümel, Fractional
  frequency collective parametric resonances of an ion cloud in a {Paul} trap,
  Phys. Rev. A 58~(1) (1998) R34--R37.
\newblock \href {http://dx.doi.org/10.1103/PhysRevA.58.R34}
  {\path{doi:10.1103/PhysRevA.58.R34}}.

\bibitem{zhao_parametric_2002}
X.~Zhao, V.~L. Ryjkov, H.~A. Schuessler, Parametric excitations of trapped ions
  in a linear rf ion trap, Phys. Rev. A 66~(6) (2002) 063414.
\newblock \href {http://dx.doi.org/10.1103/PhysRevA.66.063414}
  {\path{doi:10.1103/PhysRevA.66.063414}}.

\bibitem{gardner_precision_2014}
A.~Gardner, K.~Sheridan, W.~Groom, N.~Seymour-Smith, M.~Keller, Precision
  spectroscopy technique for dipole-allowed transitions in laser-cooled ions,
  Appl. Phys. B 117~(2) (2014) 755--762.
\newblock \href {http://dx.doi.org/10.1007/s00340-014-5891-1}
  {\path{doi:10.1007/s00340-014-5891-1}}.

\bibitem{collings_observation_2000}
n.~Collings, n.~Douglas, Observation of higher order quadrupole excitation
  frequencies in a linear ion trap, J. Am. Soc. Mass Spectrom. 11~(11) (2000)
  1016--1022.

\bibitem{bentine_molecular_2014}
E.~Bentine, C.~Foot, D.~Trypogeorgos, (py)lion an: open source wrapper of
  {LAMMPS} for the simulation of trapped ions, in preparation.

\bibitem{plimpton_fast_1995}
S.~Plimpton, Fast {Parallel} {Algorithms} for {Short}-{Range} {Molecular}
  {Dynamics}, Journal of Computational Physics 117~(1) (1995) 1--19.
\newblock \href {http://dx.doi.org/10.1006/jcph.1995.1039}
  {\path{doi:10.1006/jcph.1995.1039}}.

\bibitem{kahra_molecular_2012}
S.~Kahra, G.~Leschhorn, M.~Kowalewski, A.~Schiffrin, E.~Bothschafter, W.~Fues,
  R.~de~Vivie-Riedle, R.~Ernstorfer, F.~Krausz, R.~Kienberger, T.~Schaetz, A
  molecular conveyor belt by controlled delivery of single molecules into
  ultrashort laser pulses, Nat Phys 8~(3) (2012) 238--242.
\newblock \href {http://dx.doi.org/10.1038/nphys2214}
  {\path{doi:10.1038/nphys2214}}.

\bibitem{bell_ultracold_2009}
M.~T. Bell, T.~P. Softley, Ultracold molecules and ultracold chemistry,
  Molecular Physics 107~(2) (2009) 99--132.
\newblock \href {http://dx.doi.org/10.1080/00268970902724955}
  {\path{doi:10.1080/00268970902724955}}.

\bibitem{jelenkovic_sympathetically_2003}
B.~M. Jelenkovic, A.~S. Newbury, J.~J. Bollinger, W.~M. Itano, T.~B. Mitchell,
  Sympathetically cooled and compressed positron plasma, Phys. Rev. A 67~(6)
  (2003) 063406.
\newblock \href {http://dx.doi.org/10.1103/PhysRevA.67.063406}
  {\path{doi:10.1103/PhysRevA.67.063406}}.

\bibitem{kane_levitated_2010-1}
B.~E. Kane, Levitated spinning graphene flakes in an electric quadrupole ion
  trap, Phys. Rev. B 82~(11) (2010) 115441.
\newblock \href {http://dx.doi.org/10.1103/PhysRevB.82.115441}
  {\path{doi:10.1103/PhysRevB.82.115441}}.

\bibitem{zhao_molecular_2008}
X.~Zhao, P.~S. Krstic, Molecular dynamics simulation study on trapping ions in
  a nanoscale {Paul} trap, Nanotechnology 19~(19) (2008) 195702.
\newblock \href {http://dx.doi.org/10.1088/0957-4484/19/19/195702}
  {\path{doi:10.1088/0957-4484/19/19/195702}}.

\bibitem{broer_geometrical_1995}
H.~Broer, M.~Levi, Geometrical aspects of stability theory for {Hill}'s
  equations, Arch. Rational Mech. Anal. 131~(3) (1995) 225--240.
\newblock \href {http://dx.doi.org/10.1007/BF00382887}
  {\path{doi:10.1007/BF00382887}}.

\bibitem{noauthor_nist_nodate}
{NIST} {Handbook} of {Mathematical} {Functions} {\textbar} {Abstract} analysis
  {\textbar} {Cambridge} {University} {Press}.

\bibitem{march_quadrupole_2005}
R.~E. March, J.~F.~J. Todd, Quadrupole {Ion} {Trap} {Mass} {Spectrometry},
  {Volume} 165, {Second} {Edition}, Wiley-VCH, 2005.

\end{thebibliography}


%
%
%
\end{document}